\documentstyle[12pt,axodraw,epsfig]{article}

\def\ie{{\it i.e.}}
\def\lsim{\:\raisebox{-0.5ex}{$\stackrel{\textstyle<}{\sim}$}\:}
\def\gsim{\:\raisebox{-0.5ex}{$\stackrel{\textstyle>}{\sim}$}\:}

\begin{document}
{\flushright
\hspace{30mm} hep-ph/0312231\\
\hspace{30mm} KIAS-P03080 \\
\hspace{30mm} UCCHEP/24-03 \\
\hspace{30mm} December 2003 \\
}
\vspace{1cm}

\begin{center}
{\Large\sc {\bf Double fermiophobic Higgs boson production\\
 at the LHC and LC  }}
\vspace*{3mm}
\vspace{1cm}

{\large {A.G. Akeroyd}$^a$, {Marco A. D\'\i az}$^b$, 
{Francisco J. Pacheco}$^b$}
\vspace{1cm}

{\sl
a: Korea Institute for Advanced Study, 207-43 Cheongryangri 2-dong,\\
Dongdaemun-gu, Seoul 130-722, Republic of Korea \\
\vspace*{0.2cm}
b: Departamento de F\'\i sica, Universidad Cat\'olica de Chile,\\
Avenida Vicu\~na Mackenna 4860, Santiago, Chile 
}
\end{center}

\vspace{2cm}

\begin{abstract}
We consider the phenomenology of a 
fermiophobic Higgs boson ($h_f$) 
at the Large Hadron Collider (LHC) and a $e^+e^-$ 
Linear Collider (LC). At both machines
the standard production mechanisms which rely on the
coupling $h_fVV$ ($V=W^\pm,Z$) can be very suppressed at large
$\tan\beta$. In such cases the complementary channels 
$pp\to H^\pm h_f, A^0h_f$ and $e^+e^-\to A^0h_f$ offer promising
cross--sections. Together with the potentially large branching ratios
for $H^\pm\to h_fW^{*}$ and $A^0\to h_fZ^*$, these mechanisms 
would give rise to double $h_f$ production, leading to signatures
of $\gamma\gamma\gamma\gamma$, $\gamma\gamma VV$ and $VVVV$.

\end{abstract}

\newpage
\section{Introduction}
Neutral Higgs bosons ($h^0$) with branching ratios (BRs) very different 
to those of the Standard Model (SM) Higgs boson, $\phi^0$, can arise in  
extensions of the SM which contain an additional 
$SU(2)\times U(1)$ Higgs doublet, the ``Two Higgs Doublet
Model'' (2HDM) \cite{Gunion:1989we}. 
Assuming that each fermion type (up/down) couples 
to only one Higgs doublet \cite{Glashow:1976nt}, which
eliminates tree-level Higgs mediated flavour changing neutral currents,
leads to 4 distinct versions of the 2HDM
\cite{Barger:1989fj}. 
No compelling experimental evidence has been found for Higgs bosons. 
Experimental searches for $\phi^0$ at LEP concentrated on
the channel $\phi^0\to b\overline b$ \cite{Barate:2003sz},
while more recently \cite{Abbiendi:2000ug} searches
for Higgs bosons with large branching ratios (BRs) to lighter 
fermions and gluons (i.e. $c\overline c, \tau^+\tau^-, gg$ 
\cite{Akeroyd:1996di}) were performed.
The phenomena known as ``fermiophobia'' \cite{Weiler:1987an}
which signifies very suppressed or zero coupling to the fermions,
may arise in a particular version of the 2HDM called type I 
\cite{Haber:1979jt} or in models with Higgs triplets \cite{Georgi:1985nv}. 
Depending on its mass, a fermiophobic Higgs ($h_f$)
\cite{Barger:1992ty,Pois:1993ay,
Stange:1994ya,Diaz:1994pk,Akeroyd:1996hg,Barroso:1999bf,Brucher:1999tx}  
would decay dominantly to two photons, $h_f\to \gamma\gamma$, for 
$m_{h_f}< 95$ GeV or to two massive gauge bosons,
$h_f\to VV^{(*)}$, ($V=W^\pm,Z$) if $m_{h_f}> 95$ GeV 
\cite{Stange:1994ya,Diaz:1994pk}. 
The large BR to $\gamma\gamma$ would give a very clear experimental 
signature, and observation of such a particle would strongly 
constrain the possible choices of the underlying Higgs sector.

Experimental searches for fermiophobic Higgs bosons at LEP and Fermilab 
have been negative so far. Lower bounds of the order $m_{h_f}\ge 100$ GeV
have been obtained by the LEP collaborations OPAL\cite{Abbiendi:2002yc}, 
DELPHI\cite{Abreu:2001ib}, ALEPH\cite{Heister:2002ub}, 
and L3\cite{Achard:2002jh}, utilizing the channel 
$e^+e^-\to h_fZ$, $h_f\to \gamma\gamma$.
Only L3 \cite{Achard:2003jb} has considered $h_f\to WW^*$ decays. 
OPAL \cite{Abbiendi:2002yc} and DELPHI \cite{Abreu:2001ib} also searched 
in the channel $e^+e^-\to h_fA^0$, $h_f\to \gamma\gamma$.
From the Tevatron Run I, the limits on $m_{h_f}$ from the D0 and CDF 
collaborations are respectively 78.5 GeV \cite{Abbott:1998vv}
and 82 GeV \cite{Affolder:2001hx} at $95\%$ $c.l$, using the mechanism 
$qq'\to V^*\to h_fV$,$h_f\to \gamma\gamma$, with the dominant contribution 
coming from $V=W^\pm$. Run II will extend the coverage of $m_{h_f}$ beyond 
that of LEP \cite{Mrenna:2000qh},\cite{Landsberg:2000ht}.

All the above mass limits, however, assume that the $h_fVV$ coupling is of 
the same strength as the SM coupling $\phi^0VV$, which in general would 
not be the case for a $h_f$ in a realistic model in which the
$h_fVV$ coupling has an additional suppression from a 
mixing angle. Such a scenario would enable a very light $h_f$ 
($m_{h_f}<< 100$ GeV) to escape the searches at LEP and the 
Tevatron Run I. Therefore it is of interest
to consider other production mechanisms for $h_f$ 
which may still allow observable rates even when the $h_fVV$ 
coupling is suppressed. In a previous paper \cite{Akeroyd:2003bt}
we proposed several new production mechanisms at the Tevatron Run II.
In particular, the process $p\overline p\to H^\pm h_f$ offers 
promising rates if the masses of both $H^\pm$ and $h_f$ are less than 100 GeV.
These complementary mechanisms cover some of the region of suppressed 
coupling $h_fVV$, particularly if $m_{h_f} < 80$ GeV. However, for heavier
$m_{h_f}$ ($> 80$ GeV), detection prospects are diminished due to 
phase space suppression at the Tevatron energy.
In this paper we extend the analysis of \cite{Akeroyd:2003bt}
to consider the search potential at two future colliders, 
the Large Hadron Collider (LHC) and a 
$e^+e^-$ Linear Collider (LC). These colliders will offer significantly
improved detection prospects for $h_f$, and in case of
a $h_f$ being detected in Run II would allow a 
more precise determination of its properties.
Our work is organised as follows. In Section 2 we give a brief
introduction to fermiophobic Higgs bosons. Section 3 covers the
production of $h_f$ at the LHC and LC, while Section 4 contains
our numerical results. Conclusions are given in Section 5.
\section{Fermiophobic Higgs bosons}
The first studies of the phenomenology of $h_f$ can be found
in \cite{Barger:1992ty},\cite{Pois:1993ay}.
BRs for $h_f$ were presented in \cite{Stange:1994ya},\cite{Diaz:1994pk}
while its phenomenology at the Tevatron Run I was covered in
\cite{Stange:1994ya},\cite{Akeroyd:1996hg}.
Production at LEP2 and the impact of charged scalar loops
on BR$(h_f\to \gamma\gamma)$ (mediated via the trilinear coupling
$h_fH^+H^-$) were
studied in \cite{Barroso:1999bf},\cite{Brucher:1999tx}.


Such a particle may arise in a 2HDM in which
one $SU(2)\times U(1)$ Higgs doublet ($\Phi_2$) 
couples to all fermion types, while the other doublet ($\Phi_1$) 
does not. This model is usually called ``Type I''
\cite{Haber:1979jt}. Due to the mixing in the
CP--even neutral Higgs mass matrix (which is diagonalized by $\alpha$)
both CP--even eigenstates $h^0$ and $H^0$ can couple to the fermions.
The fermionic couplings of the lightest CP--even Higgs $h^0$ take the form 
$h^0f\overline f \sim \cos\alpha/\sin\beta$,
where $f$ is any fermion and $\beta$ is defined by $\tan\beta=v_2/v_1$
(where $v_i$ is the vacuum expectation value of the $i^{th}$ doublet).
Small values of $\cos\alpha$ would strongly suppress the fermionic 
couplings, 
and in the limit $\cos\alpha \to 0$ the coupling $h^0f\overline f$ would
vanish at tree--level, giving rise to   
fermiophobia,
\begin{center}
\vspace{-30pt} \hfill \\
\begin{picture}(120,70)(0,23) 
\DashLine(30,30)(60,30){3}
\Text(20,30)[]{$h_f$}
\ArrowLine(60,30)(90,60)
\Text(100,60)[]{$f$}
\ArrowLine(90,0)(60,30)
\Text(100,0)[]{$\overline f$}
\end{picture}  
$=\cos\alpha/\sin\beta \sim \, 0\,.$
\vspace{30pt} \hfill \\
\end{center}
\vspace{10pt}

Exact tree level fermiophobia is not stable under radiative corrections
\cite{Diaz:1994pk},\cite{Barroso:1999bf}. One can estimate what sort of
deviation from exact fermiophobia we could expect by considering as an 
example the quantum correction involving two $W$. To estimate the order 
of magnitude of this correction we approximate
\begin{center}
\vspace{-30pt} \hfill \\
\begin{picture}(120,70)(0,23) 

\DashLine(30,30)(60,30){3}
\Text(20,30)[]{$h_f$}
\Photon(60,30)(80,50){3}{4.5}
\ArrowLine(80,50)(110,60)
\Text(120,60)[]{$f$}
\Photon(60,30)(80,10){-3}{4.5}
\ArrowLine(110,0)(80,10)
\Text(120,0)[]{$\overline f$}
\ArrowLine(80,10)(80,50)

\end{picture}  
$\sim \, \frac{1}{16\pi^2}(gm_W)(\frac{g^2}{8})m_f
C(m_h^2,0,0;0,m_W^2,m_W^2)$
\vspace{30pt} \hfill \\
\end{center}
\vspace{10pt}
where $C$ is a generic triangular Veltman's function. If we approximate 
$C\sim 1/m_h^2$, expected in the limit of large Higgs mass, and compare 
this correction with the tree level vertex in the SM 
$g_{hff}\sim gm_f/2m_W$ we find,
\begin{equation}
\frac{\Delta g_{hff}}{g_{hff}}\sim \frac{g^2}{64\pi^2}
\left(\frac{m_W}{m_h}\right)^2\sim 10^{-4}
\end{equation}
for a Higgs mass twice as large as the $W$ mass. This indicates that 
tree-level fermiophobia is weakly affected by quantum corrections.

In general one would expect approximate 
fermiophobia, with some small coupling to fermions:
%
\begin{center}
\vspace{-30pt} \hfill \\
\begin{picture}(120,70)(0,23) 
\DashLine(30,30)(60,30){3}
\Text(20,30)[]{$h_f$}
\ArrowLine(60,30)(90,60)
\Text(100,60)[]{$f$}
\ArrowLine(90,0)(60,30)
\Text(100,0)[]{$\overline f$}
\GCirc(60,30){10}{0.5}
\end{picture}  
$\sim \, 0$
\vspace{30pt} \hfill \\
\end{center}
\vspace{10pt}
Of course, the correct renormalization of this vertex involves counterterms
that need to be fixed with experimental measurements. We mention two 
examples in the literature on how this counterterm can be fixed. In the
first example \cite{Brucher:1999tx} the authors set $\cos\alpha=0$ at
tree level, \ie, tree-level fermiophobia. In this case, the one-loop
contributions to the $h_ff\overline f$ coupling are finite and therefore 
the counterterm is finite as well. The finite part of the counterterm is 
also chosen to be zero, such that the renormalized coupling becomes equal 
to the sum of the finite one-loop graphs that contribute to it. In this 
scheme, $\cos\alpha=0$ means tree-level fermiophobia, and at one-loop the 
coupling $h_ff\overline f$ is not zero, although small, inducing a small 
$h_f\rightarrow b\overline b$ branching ratio as observed in their figures.

In the second example \cite{Arhrib:2003ph}, the authors are concerned
with the 1-loop fermionic decay width in the context of the 2HDM (Model~II), 
but nevertheless their results can be adapted
to the case of $h_f$ in the 2HDM (Model I). 
In \cite{Arhrib:2003ph} the counterterm for the angle $\alpha$ is chosen
such that there is no mixing between $h$ and $H$. This means that $\alpha$ 
is the mixing angle at one-loop implying that the one-loop coupling 
$h_f f\overline f$ is proportional to $\cos\alpha$, as can be seen 
from their formula for the decay width to fermions.
In this scheme, $\cos\alpha=0$ means one-loop fermiophobia, and 
therefore the definitions for $\alpha$ in the two schemes are not 
equivalent, and a relation between the two parameters $\alpha$ must be 
derived in order to compare.

%
%
The main decay modes of a fermiophobic Higgs are 
$h_f\rightarrow\gamma\gamma, W^*W^*, Z^*Z^*$.  
%
%
%
%
%
Assuming that $h_f\to \gamma\gamma$ is primarily mediated by the 
$W$ loop,
this photonic channel is dominant for $m_{h_f}\lsim 95$ GeV, 
with a BR near 100\% for $m_{h_f}\lsim 80$ GeV, decreasing to 
50\% at $m_{h_f}\approx 95$ GeV and to 1\% at $m_{h_f}\approx 145$ GeV.
In contrast, BR$(\phi^0\to \gamma\gamma)\approx 0.22\%$ is the
largest value in the SM, occurring around $m_{\phi^0}=120$ GeV.
The photonic decay mode is a particularly
robust sign of fermiophobia for $m_{h_f}\le 150$ GeV, above which 
BR$(h_f\to \gamma\gamma$) approaches the SM value. Fermiophobic
models permit the largest BRs to two photons, but
(smaller) enhancements relative to the SM BR  
are also possible in other models where a neutral Higgs boson ($h^0$) 
couples to some but not all quarks, either by choosing appropriate 
mixing angles \cite{Akeroyd:1998ui}, or as a consequence of model 
building \cite{Mrenna:2000qh}, \cite{Calmet:2000vx}.
Enhancements of BR$(h^0\to \gamma\gamma$) due to the scalar loop
contribution were studied in \cite{Arhrib:2003ph},\cite{Ginzburg:2001wj}
in the decoupling limit of the 2HDM (Model II).
In this paper we will focus on $h_f$ from the
2HDM (Model~I).

\section{$h_f$ production at LHC and LC}

In this section we consider the production of
$h_f$ at the LHC and LC, in both the standard mechanisms
(which depend on the $h_fVV$ coupling), and the complementary mechanisms
which produce $h_f$ together with another Higgs boson, and depend on the 
$h_f HV$ coupling with $H=H^\pm$ or $A^0$.
For studies of these complementary mechanisms in the context
of models without a $h_f$, see \cite{Djouadi:1999rc}.
We will present the
cross--sections as a function of $m_{h_f},\tan\beta$ and $m_{H^\pm}/m_A$.
Our analysis can be applied to two different scenarios:

\begin{itemize}

\item[{(i)}] {\it Detection} of a $h_f$ at the Tevatron Run II.
In this case the LHC and LC would provide confirmation as well as 
further studies of the $h_f$ properties.

\item[{(ii)}] {\it Non--observation}  of a $h_f$ at the Tevatron Run II.
In this case the  LHC and LC would probe a significantly larger
parameter space of $m_{h_f}$ and $\tan\beta$.

\end{itemize}

In the case of $h_f$ production at $e^+e^-$ colliders,
the complementary mechanism has been exploited at LEP
\cite{Abbiendi:2002yc},\cite{Abreu:2001ib} which 
searched for $e^+e^-\to A^0h_f$. So far, complementary mechanisms have not 
been considered at the Tevatron. As emphasized in \cite{Akeroyd:2003bt},
a more complete search strategy for $h_f$ at Hadron colliders 
would include such production processes.

\subsection{Standard mechanisms}

At the LHC there are two standard ways to produce $h_f$, for which
experimental simulations have been performed 
in the context of the SM Higgs boson ($\phi^0$). These are:

\begin{itemize}

\item[{(i)}] $pp\to W^* \to Wh_f$, $W\to l\nu$ (Higgsstrahlung)
\cite{Dubinin:1997du}

\item[{(ii)}] $pp\to qq h_f$ (Vector boson fusion) 
\cite{Rainwater:1997dg},\cite{Cranmer:2004uz}

\end{itemize}
\noindent
At a $e^+e^-$ LC one has the following mechanisms:

\begin{itemize}

\item[{(iii)}] $e^+e^-\to h_fZ$ (Higgsstrahlung) \cite{Boos:2000bz}

\item[{(iv)}] $e^+e^-\to h_f\nu\overline\nu$ ($W$ boson fusion)
\cite{Boos:2000bz}

\end{itemize}
\noindent


All the above mechanisms have been shown to be effective for
$\phi^0$ due to its substantial coupling to vector bosons
(for a recent application of the above processes to $h^0$ of the MSSM 
see \cite{Dedes:2003cg}). This is not necessarily
the case in the 2HDM, in which a $h_f$ may arise.
In the 2HDM the mechanisms (i) to (iv) for $h_f$ are all 
suppressed by $\sin^2(\beta-\alpha)$, which in the tree--level 
fermiophobic limit ($\alpha\to \pi/2$) in Model~I simplifies to:
\begin{equation}
VVh_f\sim \cos^2\beta\;\;(\equiv 1/(1+\tan^2\beta)) 
\end{equation}

\begin{figure}
\centerline{\protect\hbox{\epsfig{file=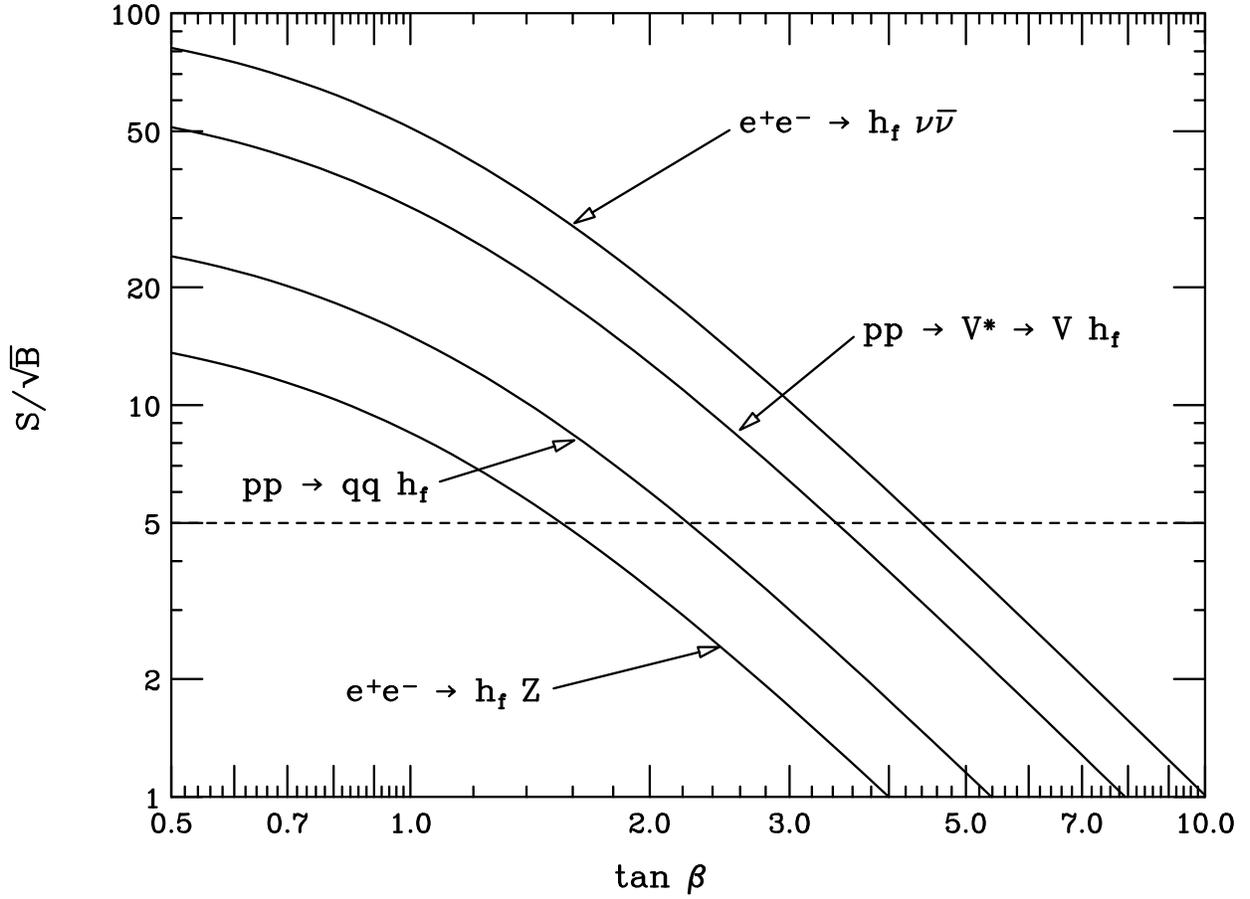,width=0.75\textwidth,angle=90}}}
\caption{\it Signal/Background ratio for a fermiophobic Higgs $h_f$ 
decaying into two photons, as a function of $\tan\beta$. Two production 
mechanisms are considered at the LHC and two at a future LC ($\sqrt s=500$
GeV).
}
\label{SrootB}
\end{figure} 
This is a severe suppression for $\tan\beta \ge 10$ and renders
all the above mechanisms unobservable (for an earlier discussion
with just mechanism (i) see \cite{Akeroyd:1998ui}).
This is shown in Fig.1, where we apply the results of the 
signal/background ($S^{\phi}/\sqrt B$) simulations
for $\phi^0\to \gamma\gamma$ to the case of a $h_f$. 
To do this we need to scale the SM Higgs signal $S^{\phi}$
by the factor BR$(h_f\to \gamma\gamma)$/BR$(\phi^0\to \gamma\gamma)$, 
and include the $\cos^2\beta$ suppression in the 
production cross--sections. Our aim is to merely show the
strong dependence of $S/\sqrt B$ on $\tan\beta$. We show results for
$S/\sqrt B > 1$ and are not concerned with additional statistical 
and systematic considerations associated with small signal rates.   
Since all the above simulations presented results for
$m_{\phi^0}=120$ GeV we will consider a $h_f$ of this mass. 
For $m_{h_f}=120$ GeV one has \cite{Stange:1994ya},
\cite{Diaz:1994pk}:
\begin{equation}
BR(h_f\to \gamma\gamma)/BR(\phi^0\to \gamma\gamma)\approx 10
\end{equation}
In Fig.1 we plot $S/\sqrt B$ for $h_f$ as a function of $\tan\beta$.
We include the production mechanisms (i)-(iv) and take $m_{h_f}=120$ GeV.
For mechanism (ii) we use the results of the simulation in
\cite{Cranmer:2004uz}.
Each curve is of the simple form: 
\begin{equation}
S/\sqrt B=10K_i\cos^2\beta
\end{equation}
where $K_i$ ($i=1,4$) corresponds to the SM Higgs
$S^{\phi}/\sqrt B$ for each of the mechanisms (i)-(iv). 
We assume luminosities (${\cal L}$) of 50 fb$^{-1}$ for (i),(ii) and 
1000 fb$^{-1}$ for (iii),(iv).
For other choices of ${\cal L}$ the $S/\sqrt B$ scales as 
$\sqrt {\cal L}$.
One can see that all the mechanisms offer spectacular signals
($S/\sqrt B>> 5$) {\sl when} there is little suppression
in the cross-section at low $\tan\beta$. However, 
$S/\sqrt B$ falls rapidly as $\tan\beta$ increases, and
$S/\sqrt B< 5$ at some critical value $\tan\beta_{C}$.
In Fig.1, $\tan\beta_{C}$ varies between 2 and 5.
Hence unless $\tan\beta$ is fairly small a relatively light $h_f$ 
(even $m_{h_f}<< 120$ GeV) may escape detection
at both the LHC and LC. 

\subsection{Complementary mechanisms}

Complementary mechanisms play an important role in the search
for $h_f$ in the case of the $h_fVV$ coupling being suppressed.
The process $p\overline p\to H^\pm h_f$ \cite{Akeroyd:2003bt}
at the Tevatron Run II, 
although offering promising rates for lighter $m_{h_f}$ is 
significantly suppressed for $m_{h_f}, m_{H^\pm} > 100$ GeV. 
We shall consider the following direct production mechanisms of $h_f$,

\begin{itemize}

\item [{(i)}] At the LHC: $pp\to H^\pm h_f,A^0h_f$ 

\item [{(ii)}] At a LC: $e^+e^-\to A^0 h_f$ 

\end{itemize}
\begin{center}
\vspace{-40pt} \hfill \\
\hspace{1cm}
\begin{picture}(200,70)(0,25) 
\Photon(60,25)(118,25){4}{8}
\ArrowLine(60,25)(10,55)
\ArrowLine(10,-5)(60,25)
\DashLine(168,-5)(118,25){3}
\DashLine(118,25)(168,55){3}
\Text(2,55)[]{$\overline q'$}
\Text(2,-5)[]{$q$}
\Text(179,-2)[]{$h_f$}
\Text(179,55)[]{$H^{\pm}$}
\Text(90,38)[]{$W^{\pm}$}
\end{picture}
\end{center}
\vspace{1cm}
\begin{center}
\vspace{-40pt} \hfill \\
\hspace{1cm}
\begin{picture}(200,70)(0,25) 
\Photon(60,25)(118,25){4}{8}
\ArrowLine(60,25)(10,55)
\ArrowLine(10,-5)(60,25)
\DashLine(168,-5)(118,25){3}
\DashLine(118,25)(168,55){3}
\Text(2,55)[]{$e^+$}
\Text(2,-5)[]{$e^-$}
\Text(179,-2)[]{$h_f$}
\Text(179,55)[]{$A^0$}
\Text(90,38)[]{$Z$}
\end{picture}
\end{center}
\vspace{1cm}

We are not aware of explicit signal--background 
simulations for these channels. Mechanism (i) is expected to
be ineffective for decays of Higgs bosons to fermions,
but for the case of $h_f\to \gamma\gamma$ might offer more 
promising detection prospects. 
Mechanism (ii) is the LC analogy of the LEP2 process,
and is usually absent in discussions of 
the MSSM Higgs bosons due to the strong suppression
of $\cos^2(\beta-\alpha)$ for $m_A\ge m_Z$ in such models. 
However for a $h_f$ in 
the region of suppressed $h_fVV$ coupling it offers promising rates.  
Detection prospects for $e^+e^-\to A^0 h_f, 
h_f\to \gamma\gamma$ at larger $\tan\beta$
might be comparable to those for the Higgsstrahlung
channel $e^+e^-\to Zh_f$ at low $\tan\beta$ (see Fig.1).

The cross-section formulae for all the 
processes can be found in \cite{Eichten:1984eu},\cite{Dawson:1998py}.
They depend on three input parameters, $m_{h_f}$, $\tan\beta$ and
one of $m_A,m_{H^\pm}$. We sum over 
$\sigma(pp\to H^+ h_f$) and $\sigma(pp\to H^- h_f$).

\subsection{Decays $H^\pm\rightarrow h_fW^*$ and $A^0\rightarrow h_fZ^*$}

The experimental signature arising from the complementary 
mechanisms in section 3.2 depends on the decay products of
$H^\pm$ and $A^0$. It has been shown \cite{Akeroyd:1998dt} (see also
\cite{Borzumati:1998xr})
that both BR($H^\pm\rightarrow h_fW^*$) and BR($A^0\rightarrow h_fZ^*$)
can be very large in the 2HDM (Model~I) since the decay widths to the
fermions ($H^\pm\to f'\overline f, A^0\to f\overline f$)
scale as $1/\tan^2\beta$. Thus in the region of $\tan\beta>10$
(where the complementary mechanisms are important) 
the fermionic channels are very suppressed, enabling the decays
$H^\pm\rightarrow h_fW^*$ and $A^0\rightarrow h_fZ^*$ to become
the dominant channels. Ref.\cite{Akeroyd:1998dt} studied the BRs
for Higgs boson masses of interest at LEP2. In this paper we
are extending their analysis to include masses of interest at the LHC and a
LC.

\begin{figure}
\centerline{\protect\hbox{\epsfig{file=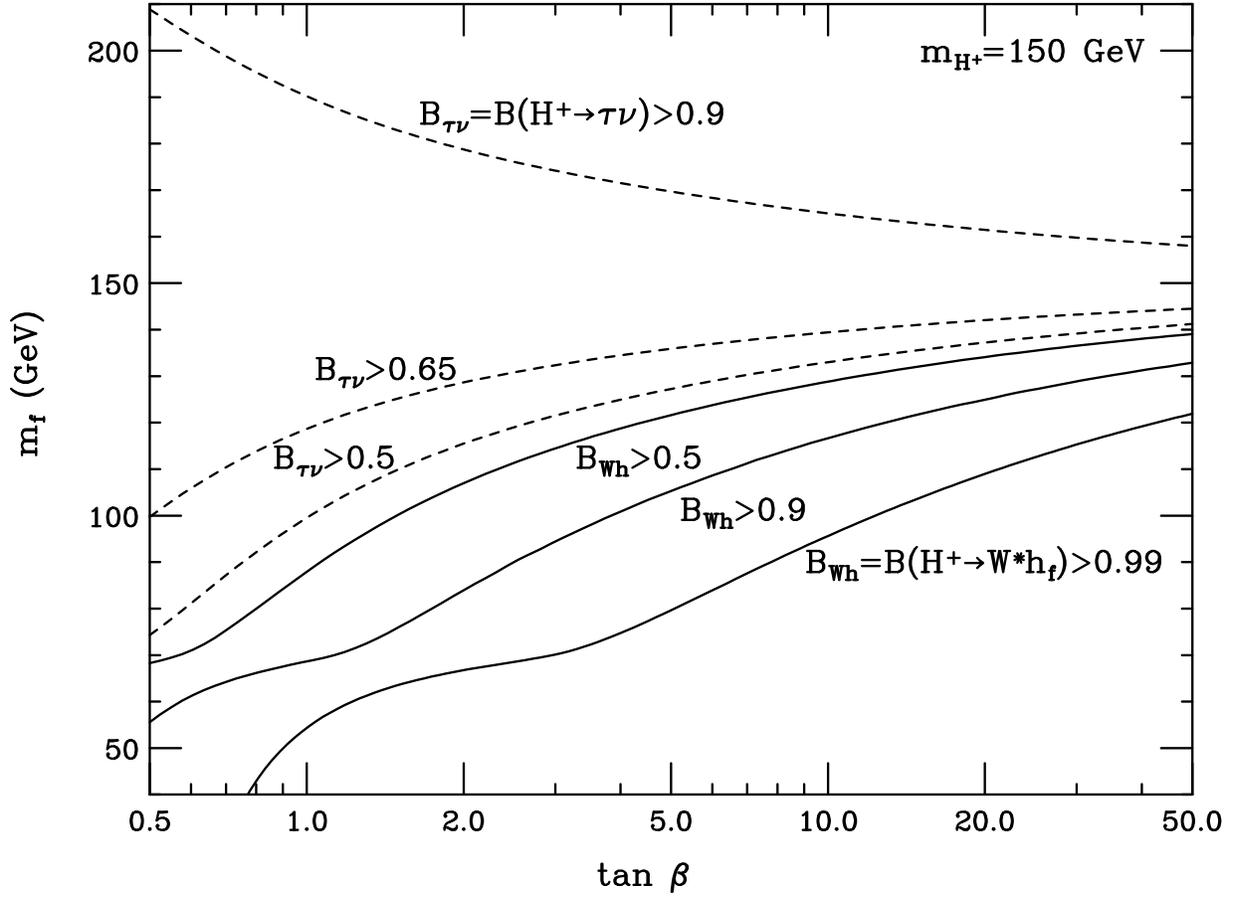,width=0.75\textwidth,angle=90}}}
\caption{\it 
Curves with constant branching ratios $BR(H^\pm\rightarrow W^*h_f)=$0.5, 0.9, 
and 0.99, and $BR(H^\pm\rightarrow\tau\nu)=0.5$, 0.65, and 0.9, in the 
$m_{h_f}-\tan\beta$ plane for $m_{H^+}=150$ GeV. 
}
\label{BRch_tb2}
\end{figure} 
\begin{figure}
\centerline{\protect\hbox{\epsfig{file=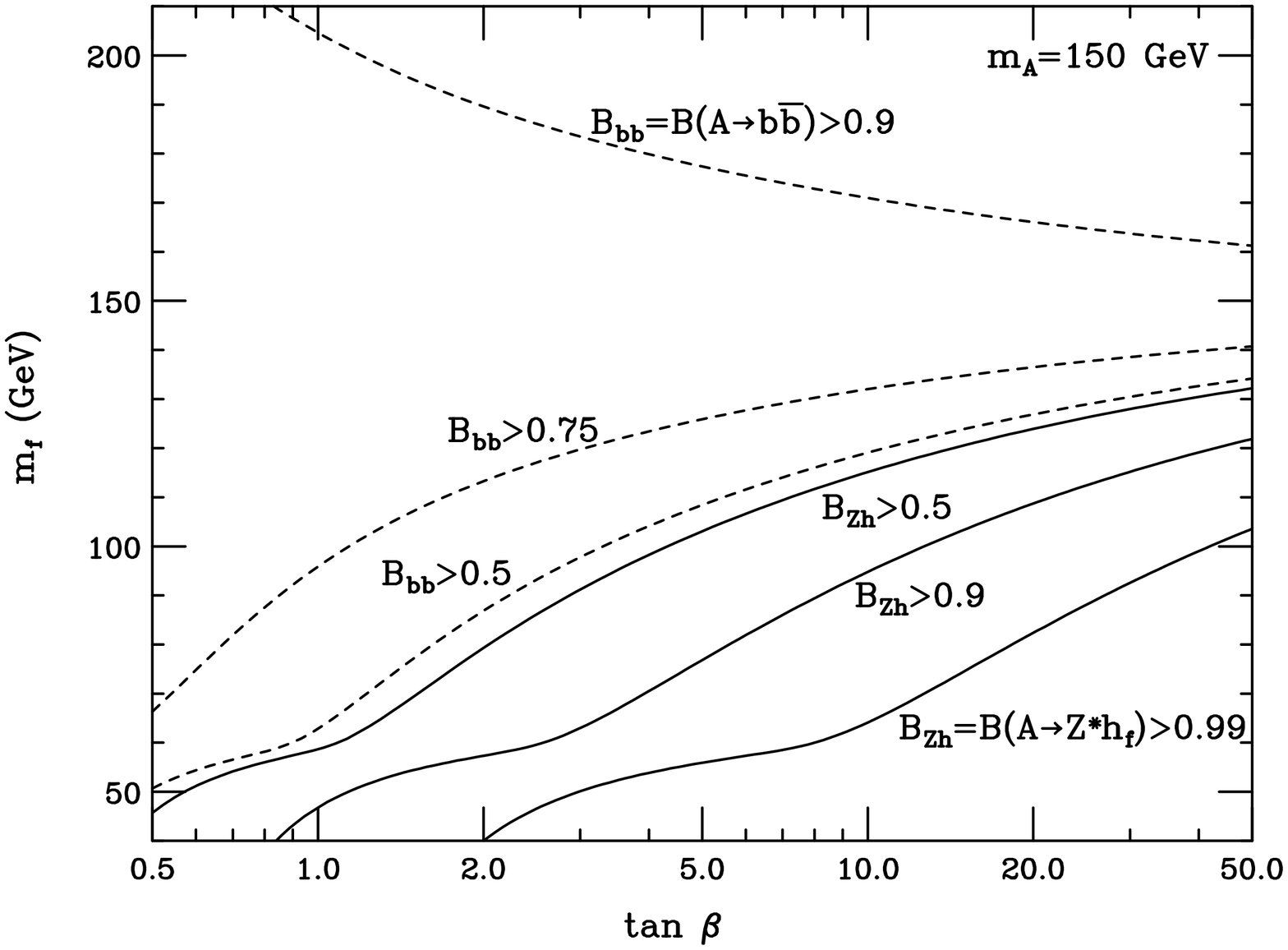,width=0.75\textwidth,angle=90}}}
\caption{\it 
Curves with constant branching ratios $BR(A^0\rightarrow Z^*h_f)=0.5$, 0.9, 
and 0.99, and $BR(A^0\rightarrow b\overline b)=0.5$, 0.75, and 0.9, in the 
$m_{h_f}-\tan\beta$ plane for $m_A=150$ GeV. 
}
\label{BRA_tb2}
\end{figure} 
%
In Fig.~\ref{BRch_tb2} we plot curves of 
constant charged Higgs branching ratio in the $m_{h_f}-\tan\beta$ plane for
$m_{H^\pm}=150$ GeV. The solid curves correspond to 
$BR(H^\pm\rightarrow W^*h_f)$ and the dashed lines correspond to
$BR(H^\pm\rightarrow\tau\nu)$. The decay that interests us here, 
$H^\pm\rightarrow W^*h_f$, dominates at low values of $m_{h_f}$ 
because in this case $W^*$ is more on-shell; it also dominates
at large values of $\tan\beta$ because the competing $H^\pm f\overline f$
decays are suppressed by $1/\tan^2\beta$.
In contrast, the decay $H^\pm\rightarrow\tau\nu$ dominates at 
large values of $m_{h_f}$ and small values of $\tan\beta$. 
For $m_{f}>150$ GeV, the fermiophobic Higgs is no longer real.

Fig.~\ref{BRA_tb2} is a similar plot where we have curves with constant
CP-odd Higgs branching ratios in the $m_{h_f}-\tan\beta$ plane for $m_A=150$ 
GeV. As in the previous figure, $BR(A^0\rightarrow Zh_f)$ is in solid
lines and dominates when $m_{h_f}$ is small and $\tan\beta$ is large, and
$BR(A^0\rightarrow b\overline b)$ is in dashed lines and dominates when $m_{h_f}$
is large and $\tan\beta$ is small. 
By comparing
Figs.~\ref{BRch_tb2} and \ref{BRA_tb2} it is apparent
that the region of domination of the
decay $A^0\rightarrow Zh_f$ in the $m_{h_f}-\tan\beta$ plane is
smaller than that for the decay  
$H^\pm\rightarrow Wh_f$. This is because the decay width for 
$A^0\rightarrow b\overline b$ is larger
than that for $H^\pm\rightarrow\tau\nu$, since the former 
$\sim m^2_b$ while the latter $\sim m^2_\tau$. 

In the lower regions of $m_{h_f}-\tan\beta$ parameter space where 
$BR(H^\pm\rightarrow W^*h_f)>0.5$ and $BR(A^0\rightarrow Zh_f)>0.5$, 
a directly
produced fermiophobic Higgs boson may be accompanied by one 
produced indirectly from the decay of $H^\pm$ or $A^0$. This scenario would 
give rise to double $h_f$ production, with subsequent decay of 
$h_fh_f\to \gamma\gamma\gamma\gamma, VV\gamma\gamma$ and $VVVV$. 
For light $h_f$  ($m_{h_f}< 80 $ GeV),  
the signal $\gamma\gamma\gamma\gamma$ would dominate,
as discussed in \cite{Akeroyd:1998dt} at LEP, and
in \cite{Akeroyd:2003bt} for the Tevatron Run II. For
$m_{h_f}\approx 95$ GeV the channels $VV\gamma\gamma$ and $VVVV$ 
would be comparable in number to $\gamma\gamma\gamma\gamma$, 
while for $m_{h_f}> 100$ GeV, the $VVVV$ would start to 
be the dominant signature. We stress that  
double $h_f$ production requires a large 
BR($H^\pm\rightarrow h_fW^*$) or BR($A^0\rightarrow h_fZ^*$) and is a feature
of the 2HDM (Model~I). The analogous BRs in other versions of the 2HDM 
are much smaller, although large BRs are also possible in triplet models
with fermiophobia \cite{Akeroyd:1998zr}.

%
%

\section{Production Cross Sections}

For the production cross--sections at the LHC
we shall be using the MRST2002 set from \cite{Martin:2001es}.
Note that QCD corrections increase the tree--level cross--section by
a factor of around 1.3 \cite{Dawson:1998py}. In our analysis we shall 
present results using the tree--level formulae only. 
In the following figures we plot contour lines of constant cross--section 
at both the LC and LHC for different choices of parameters
$m_{h_f},\tan\beta,m_{H^\pm,A^0}$. We will
show results for $e^+e^-\to h_fA^0$ and $pp\to H^\pm h_f$. The 
cross--section for $pp\to h_fA^0$ is half that of $pp\to H^\pm h_f$,
for $m_A=m_{H^\pm}$.

\begin{figure}
\centerline{\protect\hbox{\epsfig{file=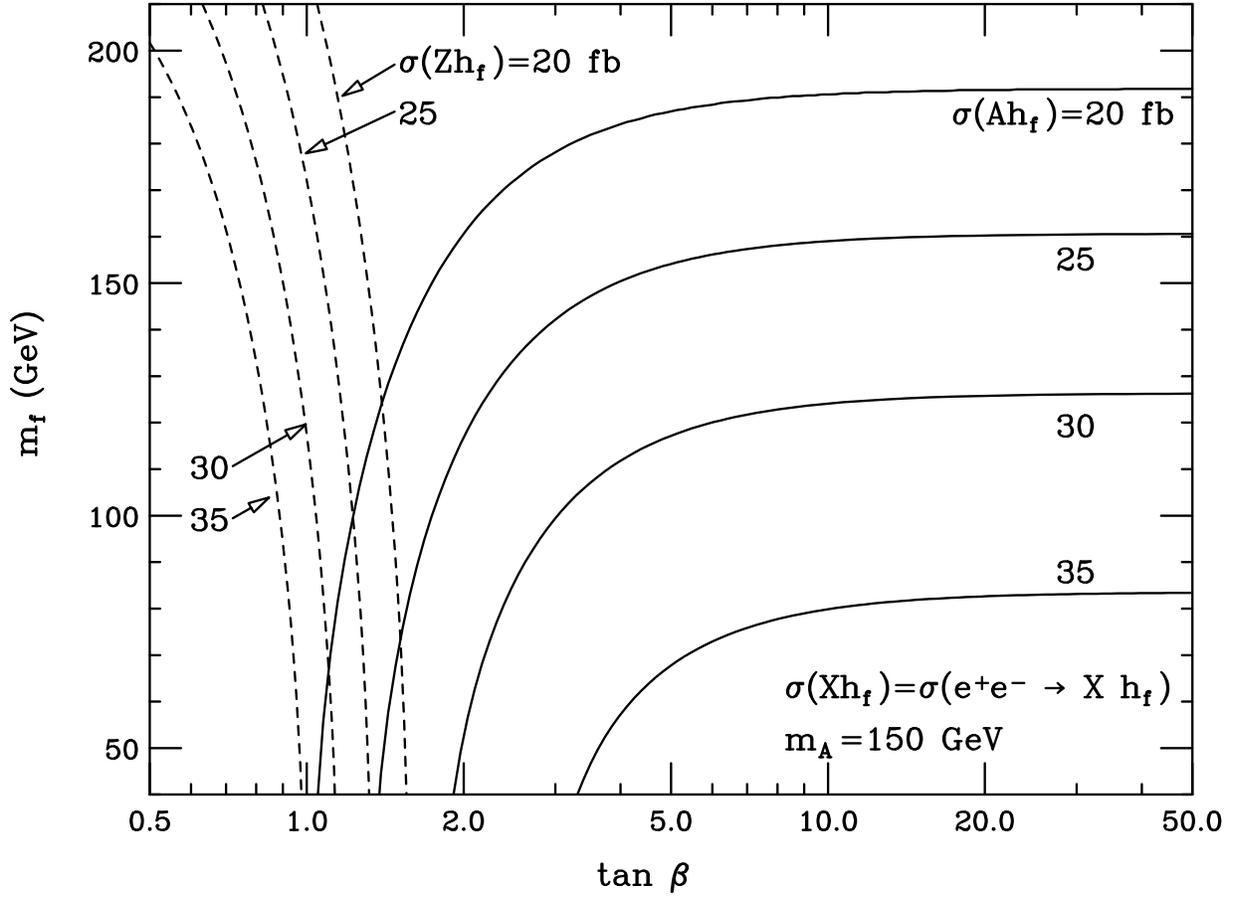,width=0.75\textwidth,angle=90}}}
\caption{\it 
Curves with constant production cross section $\sigma=20$, 25, 30, and 35 fb
at a future LC in the $m_{h_f}-\tan\beta$ plane for $m_A=150$ GeV. 
}
\label{mf_tbLC}
\end{figure} 
\begin{figure}
\centerline{\protect\hbox{\epsfig{file=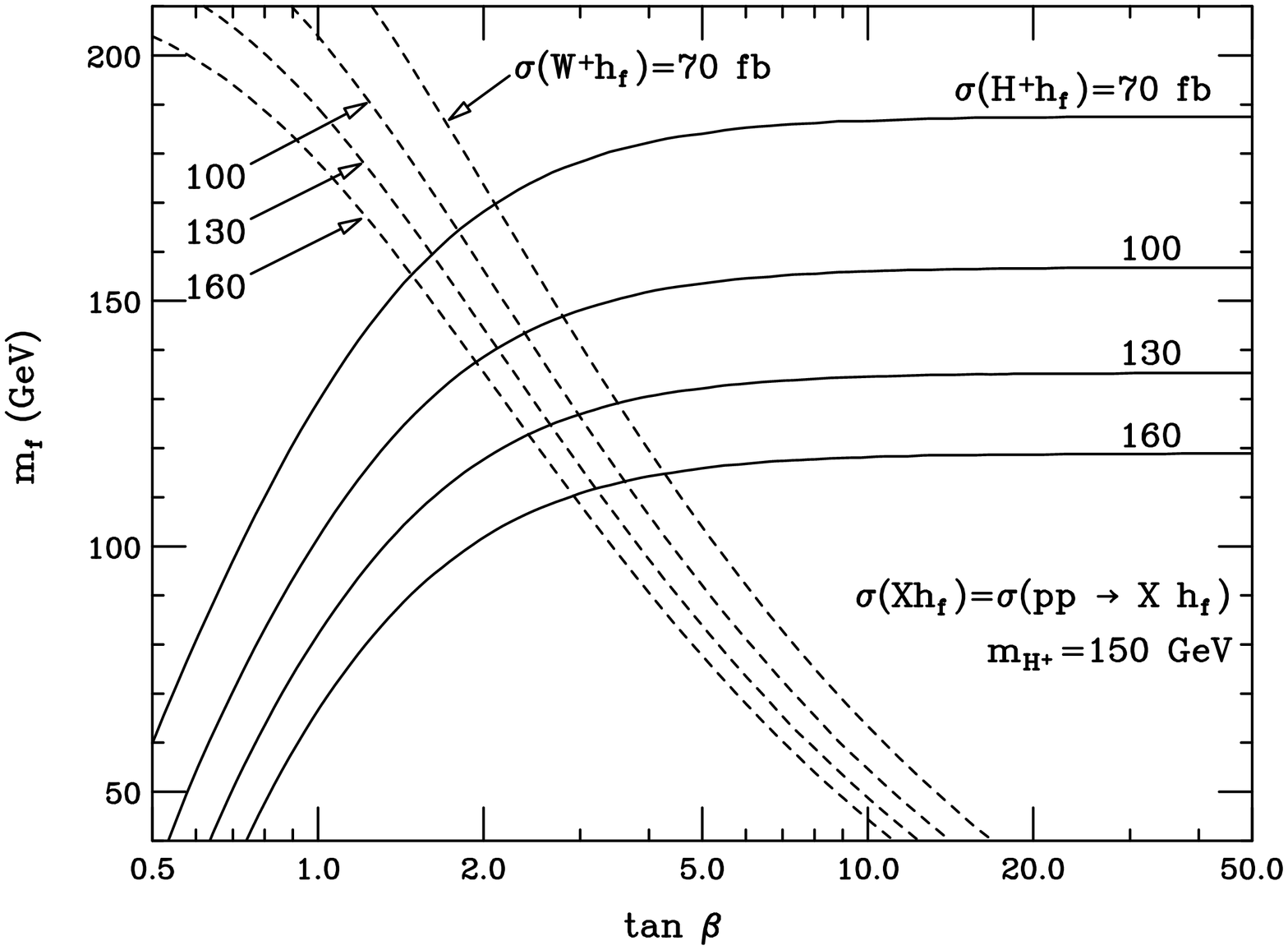,width=0.75\textwidth,angle=90}}}
\caption{\it 
Curves with constant production cross section $\sigma=70$, 100, 130, and 160 fb
at the LHC in the $m_{h_f}-\tan\beta$ plane for $m_{H^+}=150$ GeV. 
}
\label{mf_tbLHC}
\end{figure} 
In Fig.~\ref{mf_tbLC} we have contours of constant production cross section
at the LC with $\sqrt{s}=500$ GeV in the $\tan\beta-m_{h_f}$ plane, where 
$m_{h_f}$ is the fermiophobic Higgs boson mass. The four dashed lines 
correspond to the standard production mechanism $e^+e^-\rightarrow Zh_f$, 
with its cross section being equal to $\sigma(Zh_f)=20$, 25, 30, and 35 fb. 
The four solid lines correspond to the complementary mechanism 
$e^+e^-\rightarrow A^0h_f$ with the same values for its cross section 
$\sigma(A^0h_f)$, and taking $m_A=150$ GeV. The higgsstrahlung production 
mechanism dominates at small $\tan\beta$ since, in this model, the cross 
section is proportional to $\cos^2\beta$ (as explained in Section 3.1).
On the contrary, the production 
of a fermiophobic Higgs in association with a CP-odd Higgs $A^0$ dominates at 
large $\tan\beta$ due to the dependence of the cross section on 
$\sin^2\beta$. For this reason, in the case of $\sigma(Zh_f)$ the constant 
cross section contours strongly depend on $\tan\beta$, and for 
$\tan\beta\gsim 2$ the cross section is already smaller than 20 fb. Equally 
sharp but opposite dependence on $\tan\beta$ is observed for the constant 
$\sigma(A^0h_f)$ contours. This effect is evident as a clear depression 
of the observability of $h_f$ at around $\tan\beta=$1-2, where 
both cross sections are smaller than 20 fb for $m_{h_f}\gsim 130$ GeV.

In Fig.~\ref{mf_tbLHC} we have similar contours of constant production 
cross section, but this time at the LHC with $\sqrt{s}=14$ TeV, in the 
$\tan\beta-m_{h_f}$ plane. The four dashed lines correspond to the standard
mechanism $pp\rightarrow Wh_f$ with values $\sigma(Wh_f)=70$, 100, 130, and 
160 fb. The four solid lines correspond to the complementary mechanism
$pp\rightarrow H^\pm h_f$  for the same values of the cross section 
$\sigma(H^\pm h_f)$, and taking $m_{H^\pm}=150$ GeV. As before, the standard
mechanism is dominant at low values of $\tan\beta$ and the complementary
mechanism dominates at high $\tan\beta$ and this is due to a dependence 
of the partonic cross section on $\cos^2\beta$ and $\sin^2\beta$ 
respectively. Due to phase space effects, the dependence on $m_{h_f}$ is 
stronger compared with the LC case, making the equal cross section contours 
less vertical. For this reason, the depression already observed in the 
previous figure is less pronounced at the LHC.

\begin{figure}
\centerline{\protect\hbox{\epsfig{file=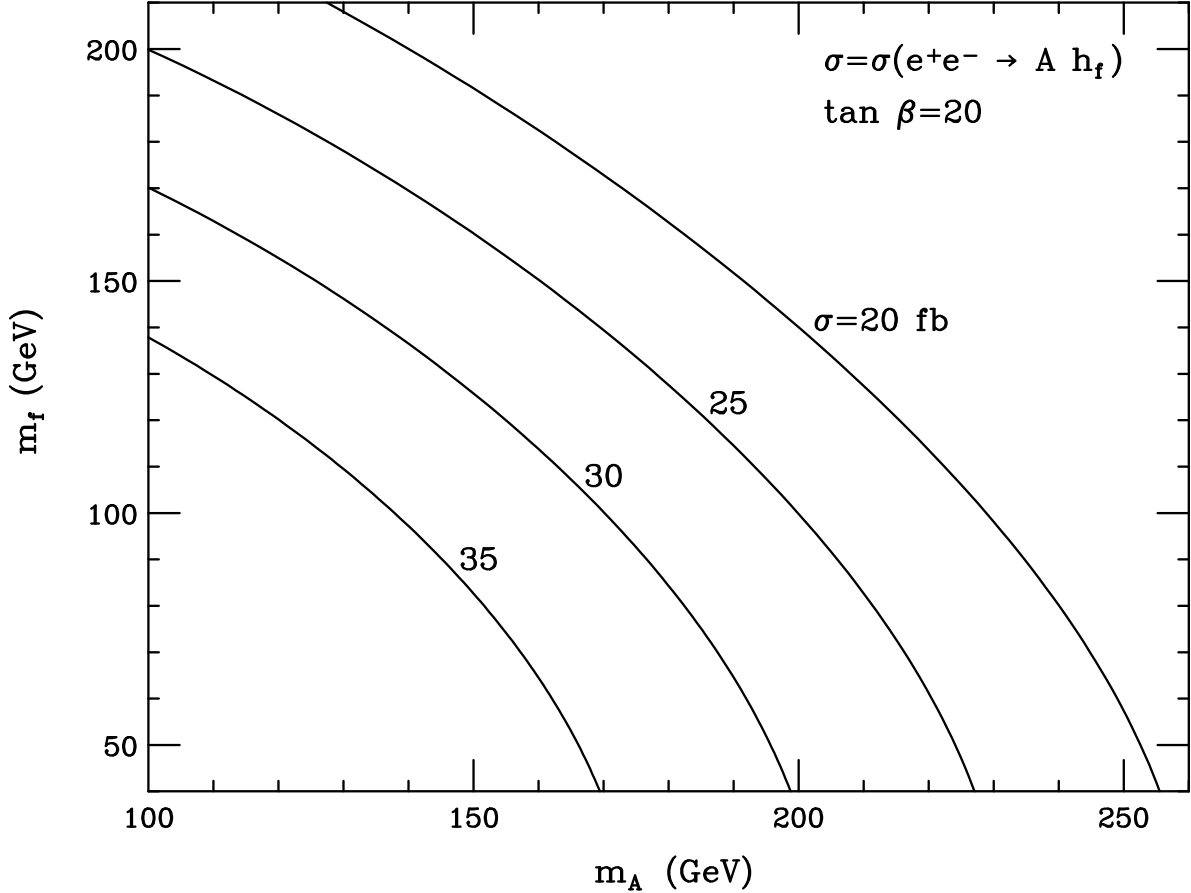,width=0.75\textwidth,angle=90}}}
\caption{\it 
Curves with constant production cross section $\sigma=20$, 25, 30, and 35 fb
at a future LC in the $m_{h_f}-m_A$ plane for $\tan\beta=20$. 
}
\label{mf_maLC}
\end{figure} 
\begin{figure}
\centerline{\protect\hbox{\epsfig{file=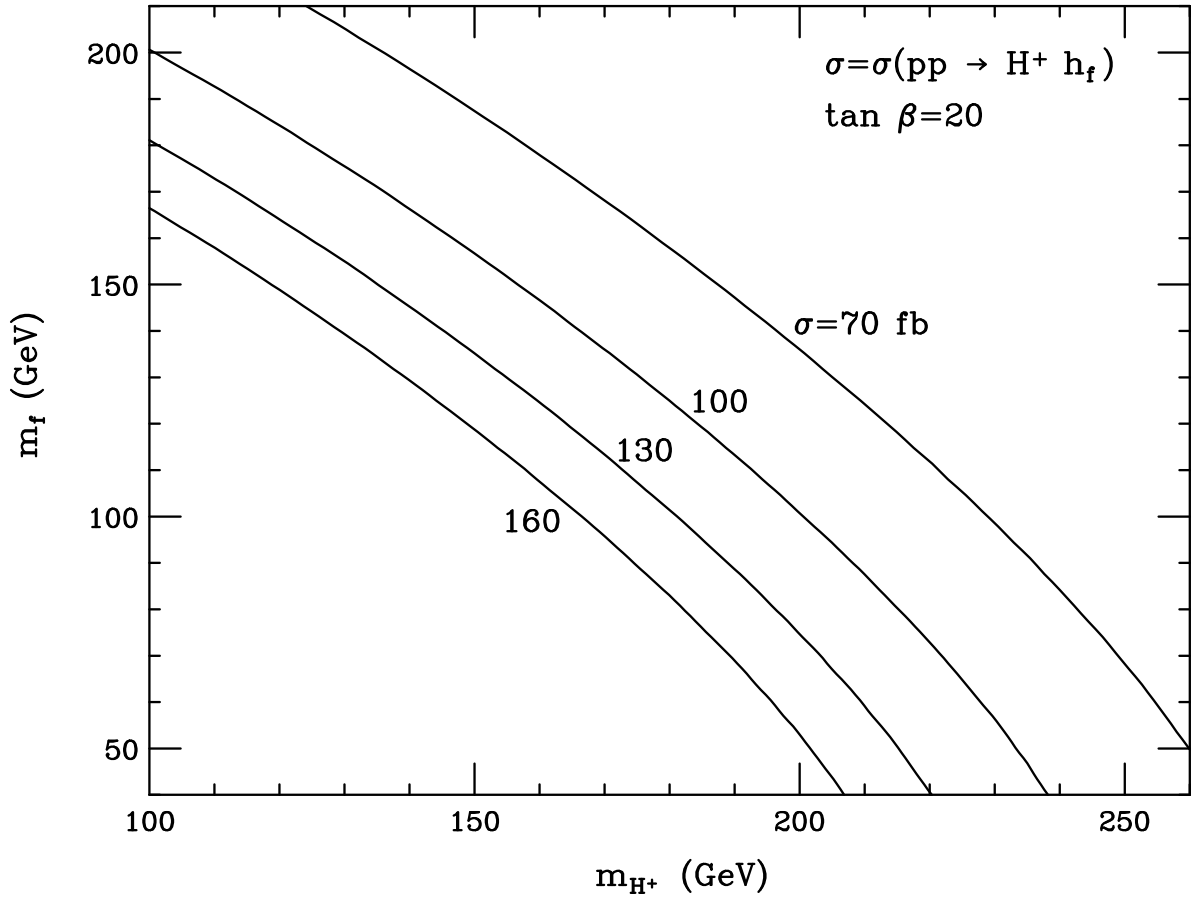,width=0.75\textwidth,angle=90}}}
\caption{\it 
Curves with constant production cross section $\sigma=70$, 100, 130, and 160 fb
at the LHC in the $m_{h_f}-m_{H^+}$ plane for $\tan\beta=20$ GeV. 
}
\label{mf_mchLHC}
\end{figure} 
In Fig.~\ref{mf_maLC} and \ref{mf_mchLHC} we plot contours of constant 
production cross section at the LC and LHC respectively, in the 
$m_A-m_{h_f}$ plane for the LC and in the $m_{H^+}-m_{h_f}$ plane for the LHC, 
using the same numerical values for the cross sections as in 
Figs.~\ref{mf_tbLC} and \ref{mf_tbLHC}. In both cases we take 
$\tan\beta=20$, where the standard production mechanisms are very 
suppressed. If a realistic simulation of the signal were made and the minimum 
number of events $N_{min}$ were known for the signal to be observable, the 
observable cross sections would be of the type 
$\sigma>N_{min}/{\cal L}$, implying that the region below and to the left
of the curves in both figures would be observable. From the figures we see 
that to increase the region of observability, the minimum cross section 
needs to be decreased more sharply at the LHC rather than at the LC.

\begin{figure}
\centerline{\protect\hbox{\epsfig{file=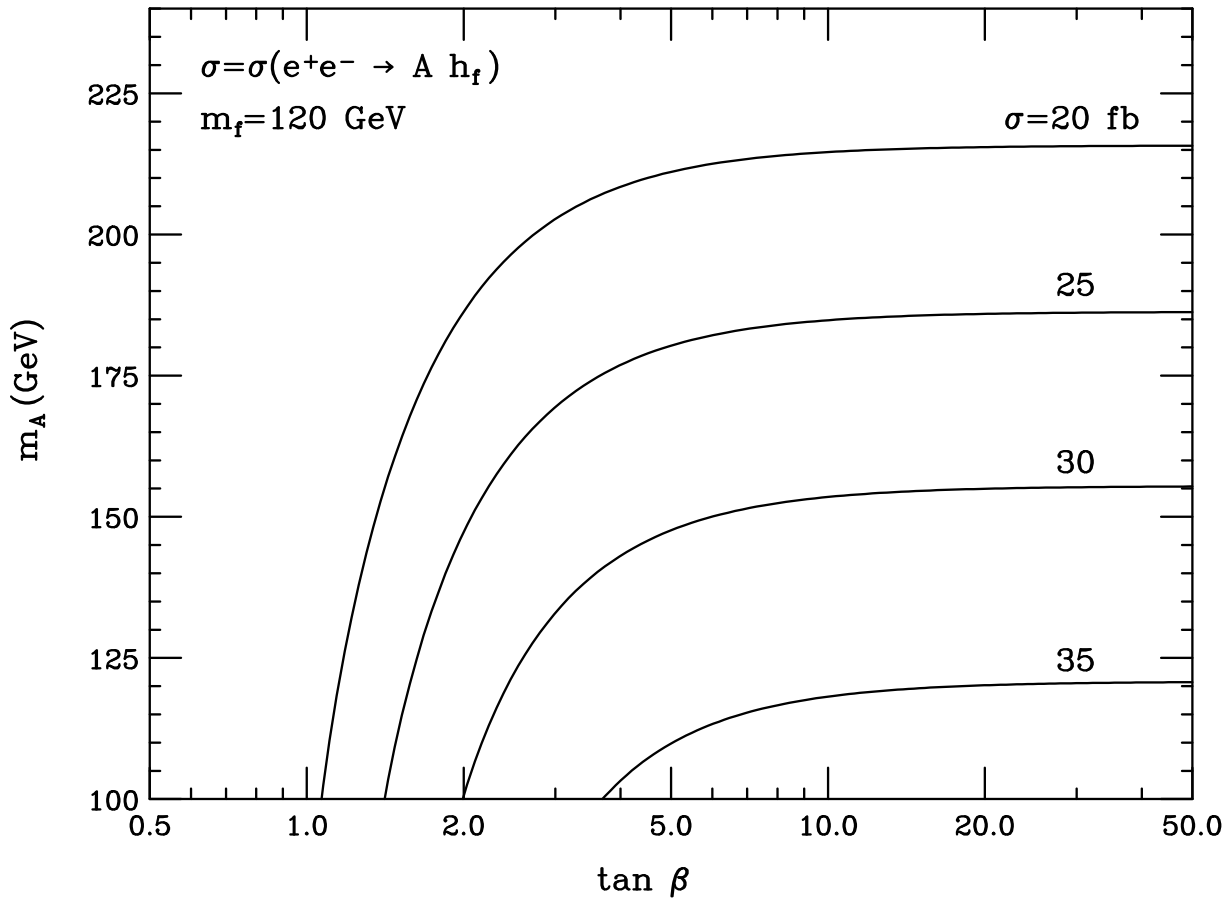,width=0.75\textwidth,angle=90}}}
\caption{\it 
Curves with constant production cross section $\sigma=20$, 25, 30, and 35 fb
at a future LC in the $m_A-\tan\beta$ plane for $m_{h_f}=120$ GeV. 
}
\label{ma_tbLC}
\end{figure} 
\begin{figure}
\centerline{\protect\hbox{\epsfig{file=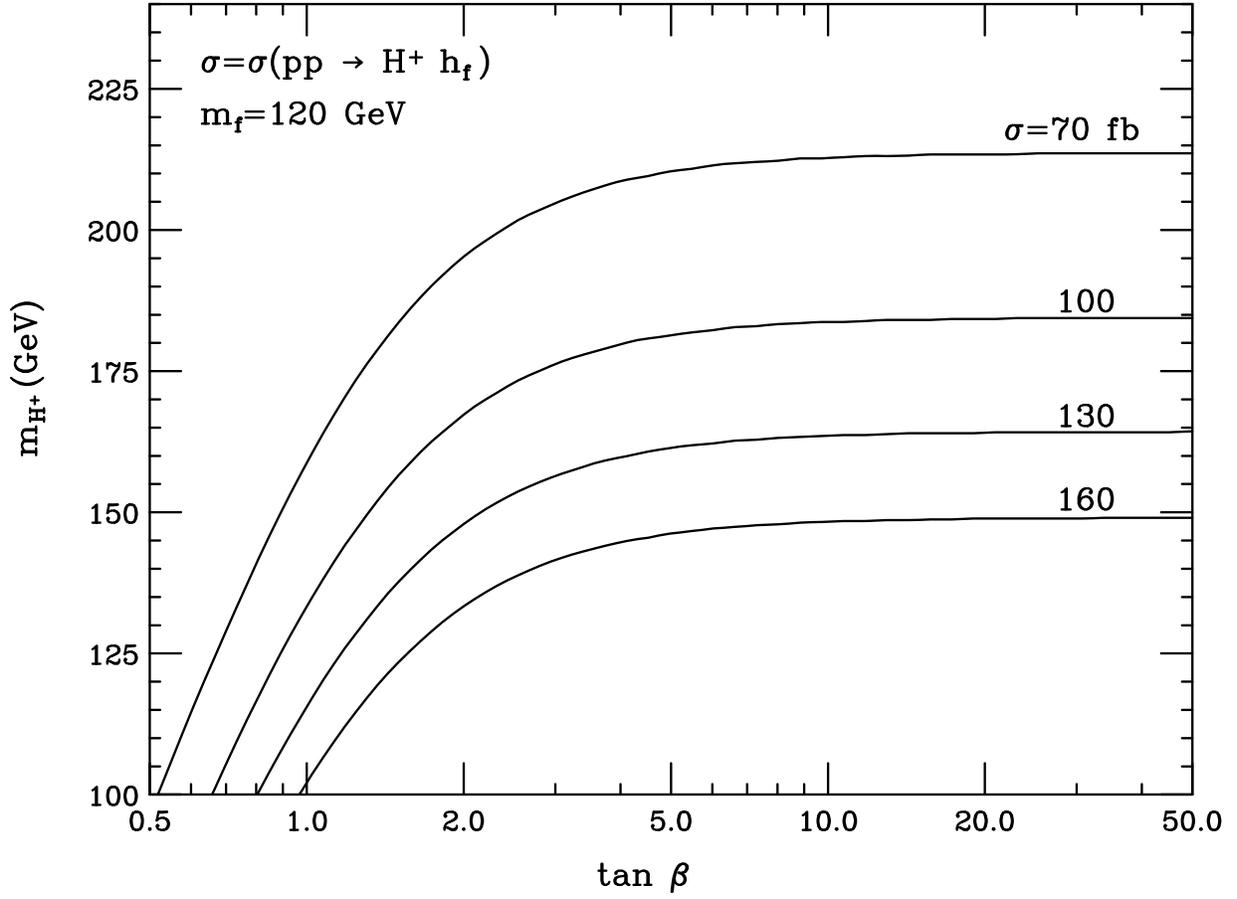,width=0.75\textwidth,angle=90}}}
\caption{\it 
Curves with constant production cross section $\sigma=70$, 100, 130, and 160 fb
at the LHC in the $m_{H^+}-\tan\beta$ plane for $m_{h_f}=120$ GeV. 
}
\label{mch_tbLHC}
\end{figure} 
In a similar way, in Figs.~\ref{ma_tbLC} and \ref{mch_tbLHC} we plot 
contours of constant production cross section at the LC and LHC respectively,
in the $m_A-\tan\beta$ plane for the LC and in the $m_{H^+}-\tan\beta$
plane for the LHC. The chosen values for the cross sections are the same as 
in the previous figures, and we take a fermiophobic Higgs mass $m_{h_f}=120$
GeV in both cases. The largest cross sections lie towards the bottom
right-hand corner of the figure. 
The stronger dependence of $\sigma(pp\rightarrow H^\pm h_f)$ on 
$m_{H^+}$ is evident from the figures when compared with the dependence
of $\sigma(e^+e^-\rightarrow A^0h_f)$ on $m_A$. 

In all the situations studied here the directly produced fermiophobic Higgs
boson decays into two photons with a branching ratio close to unity if 
$m_{h_f}\lsim 80$ GeV, close to 0.5 for $m_{h_f}\sim 95$ GeV, and near 0.01 for
$m_{h_f}\sim 145$ GeV. In the case of complementary production at the LHC and
LC shown in the previous graphs, the number of 4 photon events
will be maximized for larger $BR(A^0\rightarrow Zh_f)$ and 
$BR(H^\pm\rightarrow W^*h_f)$ and lower $m_{h_f}$.
Comparing Figs.~\ref{BRA_tb2} and \ref{mf_tbLC} we can see that (for
$m_A=150$ GeV) if the model lies below the curve 
$\sigma(e^+e^-\rightarrow A^0h_f)\approx30$ fb in the $m_{h_f}-\tan\beta$ plane
then the majority of events will be of the four photon type at 
the LC. Similarly, comparing
Figs.~\ref{BRch_tb2} and \ref{mf_tbLHC} we see that (for $m_{H^+}=150$ GeV)
if the model lies below the curve $\sigma(pp\rightarrow H^\pm h_f)
\approx 130$ fb
in the $m_{h_f}-\tan\beta$ plane, then a four photon signal would be 
plentiful at the LHC. Identifying a lepton from the decay of $W^*$ or 
$Z^*$ would further reduce backgrounds. 
After applying a realistic
photon identification efficiency for 4 photons ($0.8^4\approx 0.4$)
\cite{Cranmer:2004uz} and multiplying by the appropriate BR factors,
signal sizes for $\gamma\gamma\gamma\gamma +l^\pm$
in excess of a few fb are possible in a sizeable
region of the $m_{h_f}-\tan\beta$ plane.

At the LHC, the main backgrounds for the $\gamma\gamma\gamma\gamma+ l^\pm$ 
are expected to be
the irreducible $\gamma\gamma\gamma\gamma+ l^\pm$ from genuine photon
production, and the reducible 4 jet plus $l^\pm$, where all four
jets are misidentified as a photon. MadEvent \cite{Maltoni:2002qb} 
estimates the irreducible background to be $\sim 10^{-6}$ fb and thus
entirely negligible. The 4 jet plus $l^\pm$ was estimated to be
$\sim 130,000 fb$, but this fake photon background can be reduced to a 
negligible size after applying realistic rejection factors of $10^3$
for each jet \cite{Cranmer:2004uz}. Hence we conclude that 
$\gamma\gamma\gamma\gamma + l^\pm$ is a robust, relatively background
free signature at the LHC. A detailed study of detection prospects
in all the channels 
$\gamma\gamma\gamma\gamma$, $\gamma\gamma VV$ and $VVVV$
will be considered in a separate work.

\section{Conclusions}

We have considered the phenomenology of a fermiophobic Higgs boson
$(h_f)$ at the Large Hadron Collider (LHC) 
and a $e^+e^-$ Linear Collider (LC). We showed that the 
production mechanisms $pp\to H^\pm h_f, A^0 h_f$ and $e^+e^-\to A^0h_f$
offer promising cross--sections in the region where the conventional
mechanisms $pp\to W^\pm h_f$ and $e^+e^-\to h_f Z$ are very suppressed.
A more complete search strategy at both these colliders would include these
complementary production mechanisms. The potentially large branching ratios
for $H^\pm\to h_fW^{*}$ and $A^0\to h_fZ^*$, would 
lead to double $h_f$ production, with subsequent signatures
$\gamma\gamma\gamma\gamma$, $\gamma\gamma VV$ and $VVVV$, which need 
experimental simulations. Production cross sections are similar at both 
machines, but the larger luminosity and smaller backgrounds
at the LC would permit precision 
measurements necessary to determine the exact nature of the observed 
fermiophobic Higgs boson.

\section*{Acknowledgements}  

M.A.D. is thankful to Korea Institute for Advanced Study where part of 
this work was carried out. This research was partly funded by CONICYT 
grant No.~1030948.

\end{document}